# REAL-TIME BRAIN MACHINE INTERACTION VIA SOCIAL ROBOT GESTURE CONTROL

**Reza Abiri**
Department of Mechanical, Aerospace, and Biomedical Engineering
University of Tennessee, Knoxville
Knoxville, TN 37996, USA
rabiri@vols.utk.edu

**Soheil Borhani**
Department of Mechanical, Aerospace, and Biomedical Engineering
University of Tennessee, Knoxville
Knoxville, TN 37996, USA
sborhani@vols.utk.edu

**Xiaopeng Zhao**
Department of Mechanical, Aerospace, and Biomedical Engineering
University of Tennessee, Knoxville
Knoxville, TN 37996, USA
xzhao9@utk.edu

**Yang Jiang**
Department of Behavioral Science
Sanders-Brown Center on Aging
University of Kentucky, College of Medicine
Lexington, KY 40356, USA
yjiang@uky.edu

**ABSTRACT**
Brain-Machine Interaction (BMI) system motivates interesting and promising results in forward/feedback control consistent with human intention. It holds great promise for advancements in patient care and applications to neurorehabilitation. Here, we propose a novel neurofeedback-based BCI robotic platform using a personalized social robot in order to assist patients having cognitive deficits through bilateral rehabilitation and mental training. For initial testing of the platform, electroencephalography (EEG) brainwaves of a human user were collected in real time during tasks of imaginary movements. First, the brainwaves associated with imagined body kinematics parameters were decoded to control a cursor on a computer screen in training protocol. Then, the experienced subject was able to interact with a social robot via our real-time BMI robotic platform. Corresponding to subject's imagery performance, he/she received specific gesture movements and eye color changes as neural-based feedback from the robot. This hands-free neurofeedback interaction not only can be used for mind control of a social robot's movements, but also sets the stage for application to enhancing and recovering mental abilities such as attention via training in humans by providing real-time neurofeedback from a social robot.

**INTRODUCTION**
Interest in neural-based human-robot platforms has dramatically increased during the past decades in the field of neurorehabilitation and Brain Computer Interface (BCI) [1]. By using invasive approaches in primates [2-9] and humans [10-15], some interesting BMI robotic platforms were developed. During invasive BCI, the human had electrodes surgically implanted inside or on the surface of his/her brain [6, 11]. However, due to invasiveness, this method is not a suitable solution for short-term rehabilitation programs in humans.

Recent advances in noninvasive approaches and BMI development confirmed the potential for rehabilitation using direct neural modulation related to the patient's intention in a neural-based rehabilitation robotic platform [16-22]. Currently there is a renaissance of interest in using noninvasive electroencephalography (EEG) brainwaves as a common, popular, affordable and portable method, which has been applied by many researchers to control various external devices in different paradigms [1]. EEG records synchronized neural activity of the brain within milliseconds. The recorded electrical signals from the scalp of the brain differ during resting or during different tasks, e.g. whether subjects are imaging moving the left arm or the right leg, or memorizing an image. Importantly, EEG signals, reflecting synaptic functions, differ amongst people of different ages and mental health status, e.g. between a healthy, cognitively normal (NC) person and a patient who has mild cognitive impairment (MCI) or Alzheimer's Disease (AD), when performing the same task [23, 24]. The high discriminative and noninvasive nature of the EEG signal makes it a promising candidate for brain training via BCI in persons with cognitive deficit in a neurofeedback rehabilitation platform. However, this line of research is still in its infancy.

Using noninvasive EEG monitoring, two major motor imagery paradigms have been developed in previous studies including "sensorimotor rhythms" and "imagined body kinematics." In the sensorimotor rhythms paradigm, the EEG signals are typically collected by imaginary movement of large body parts [25] and are employed to control a cursor [26-29] or various external robotic devices [30-37]. For example, Bouyarmane et al. [38] controlled the position of the moving foot in an autonomous humanoid robot with 36-degree-of-freedom based on motor imagery of arm movements. However, BCI or BMI systems based on sensorimotor rhythms require lengthy training time (some weeks to several months) to gain satisfactory performance. Recently, Bradberry et al. [39] proposed a new EEG-based BCI paradigm (natural imaginary movement) in time-domain, which can significantly reduce the training time in BCI. Using this "imagined body kinematics" paradigm, they reported positive performance in a cursor control problem after only about 40 minutes of training and practice. It is worth pointing out that previous work on invasive devices show that the subjects with implanted electrodes in brain could quickly gain high success rate in target acquisition based on continuous imagined kinematics of just one body part [6, 11].

In the current study, we aim to develop and test the feasibility of a novel BMI platform using brain signals-controlled gestures of a social robot to provide neurofeedback to a human subject. The present work is novel in providing specific neurofeedback to the subject in the context of brain training. In contrast, previously developed humanoid robot-based BMI platforms [38, 40-44] mostly investigated the direct control of robots in manipulation tasks. We hypothesized that neural-based feedback from a social robot may be more engaging and effective in maintaining user interest in specific mental tasks in the targeted rehabilitation program. An initial test of the developed platform is conducted using imagined body kinematics. The neurofeedback-based BMI platform developed here will set the foundations for the next stage of application to rehabilitation via brain training such as enhancing cognitive abilities, e.g., attention and memory, by providing real-time neurofeedback from a social robot.

**Materials and methods**

This section discusses the subject training protocol (2.1) and the brain-robot interaction platform (2.2). The Institutional Review Board at the University of Tennessee approved the experimental procedure. Five subjects participated in the experiments with written consent. None of the subjects had previously participated in BCI research studies. During the experiments, EEG signals were acquired using a wireless headset (Emotiv EPOC [45] device with 14 channels and through BCI2000 software [46](with 128Hz sampling frequency, high pass filter at 0.16Hz, and low pass filter at 30Hz).

*Subject training protocol*

Four healthy male subjects and one healthy female subject (right-handed, averaged age 24) participated in the training protocol after giving informed consent. Before direct interaction with the social robot, the naive subjects became familiar with the BCI systems via training tasks.

*Training Tasks*

The computerized task was provided by a PC with dual monitors. One monitor was viewed by the experimenter and the other by the subject.

The protocol had three phases. Phase 1 was the training phase of the hands-free cursor task. The subject was asked to sit comfortably in a chair with hands resting on the laps. The subject's face was kept at an arm's length from the monitor. The subject was instructed to track the movement (right-left/up-down) of a computer cursor, whose movements were controlled by an experimenter in a random manner. Meanwhile, the subject was taught to imagine the same matching movement velocity with their right index fingers. The training phase consisted of 5 trials for right-left and 5 trials for up-down; each trial lasted 60 seconds. Phase 2 was the calibration phase, during which a decoder model was constructed to model the velocity of the cursor as a function of the EEG waves of the subject. Based on a previous study [39], for more accurate reconstruction and prediction of the imagined kinematics at each point, 5 previous points (time lag) of EEG data (for a subject) were also included in the decoding procedure. Then, the developed decoder was fed into BCI2000 software to test the performance of the subject in phase 3 (test phase). In the test phase, the subject was asked to move the cursor using their imagination to the specific targets that randomly appeared on the center or at the edges of the monitor.

*Decoding brain signals*

Many decoding methods for EEG data have been investigated by researchers in frequency and time domains. Most of sensorimotor-rhythms-based studies were developed in the frequency domain [26-32, 34, 35, 47]. Meanwhile, in the time domain, researchers employed regression models as a common method for decoding EEG data for offline decoding [48-52] and real-time implementation [39]. Kalman filter [53] and particle filter models [54] were applied in decoding EEG signals for offline analysis, as well. Many previous works confirmed that among kinematics parameters (position, velocity), velocity encoding/decoding showed the most promising and satisfactory validation in prediction [48, 49, 51]. Hence, we were motivated to decode and map the acquired EEG data to the observed cursor velocities in x and y directions. In other words, the aim was to reconstruct the subject's imagined trajectories from EEG data and obtain a calibrated decoder. For this purpose, all the collected data were transferred to MATLAB software [55] for analyzing and developing a decoder. Here, based on a regression model for output velocities at time $t$ in x direction ($u[t]$) and y direction ($v[t]$), the equations are presented as follows:

$$u[t] = a_{0x} + \sum_{n=1}^{N}\sum_{k=0}^{K} b_{nkx} e_n[t-k] \quad (1)$$

$$v[t] = a_{0y} + \sum_{n=1}^{N}\sum_{k=0}^{K} b_{nky} e_n[t-k] \quad (2)$$

where $e_n[t-k]$ is the measured voltage for EEG channel $n$ at time t-k. The total number of EEG channels is $N = 14$ and the total lag number is chosen to be $K = 5$. The choice of 5 lag points is the tradeoff between accuracy and computational efficiency. Meanwhile, $a$ and $b$ were the parameters that could be calculated by feeding the data to the equations 1 & 2 and by MATLAB coding.

The data collected in training sessions were fed to equations 1 and 2 without any further filtering and the final developed decoder was employed to test and control the cursor on the monitor. Figure 1 shows a simple schematic of our experimental setup during the training protocol.

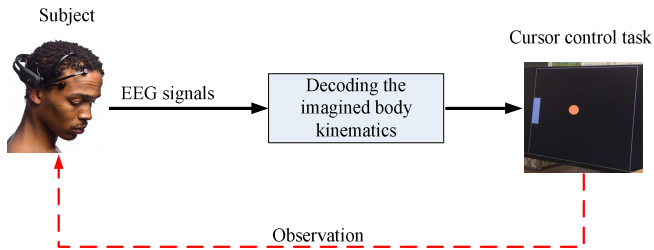

**Fig. 1:** A schematic of our EEG-based BCI platform for brain training and cursor control task

*BMI robotic platform*

This subsection discusses details of the hardware and software employed in our brain-robot interaction platform. Figure 2 shows the completed and developed BMI robotic platform with different components and the overall schematics of data flow from the brain to the social robot. By programing in BCI2000 software and MATLAB (Simulink) software [55], imaginary movement tasks (from the subject) were mapped to specific real movements and change of eyes color in a social robot and the subject could see the result of his/her imaginary performance as neurofeedback in real time and from the robot. For example, as a simple scenario, right imaginary movement was mapped to right hand movement of robot and changing eyes color to green; left imaginary movement was mapped to left hand movement of the robot and changing eyes color to blue. The proposed platform as a total system operated as a closed-loop system with bilateral interaction. A brief description for each part of the platform is presented below.

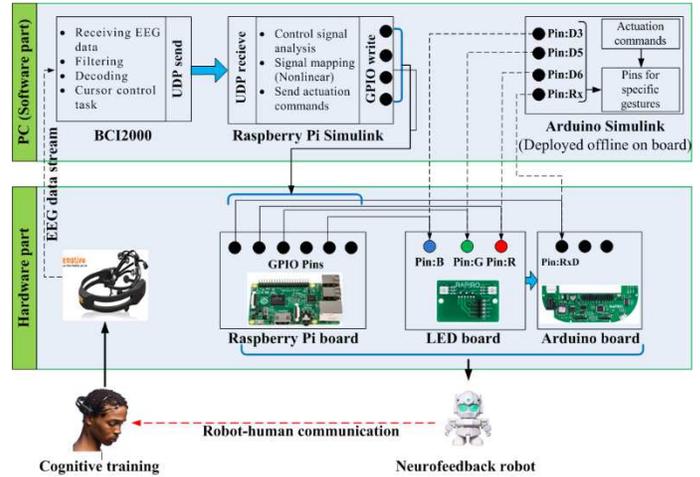

**Fig. 2:** The different components of the developed real-time neurofeedback-based BMI system

*Interface software*

All collected EEG signals were transferred wirelessly to BCI2000 where the EEG signals were used in the decoding algorithm to compute intended velocities. The generated velocity signals were sent in real time to the cursor control application module and to Raspberry Pi Simulink via User Datagram Protocol (UDP) [56]. As shown in Figure 2, the hardware part of the platform (Raspberry Pi board, LED board, Arduino board) was modified and designed in a way that the control signals (in Simulink) could be encoded to activate specific corresponding gesture and eyes color in the social robot. Raspberry Pi Simulink was developed to communicate with the Raspberry Pi board which could control the social robot based on received signals and sent through an Arduino board. The Arduino Simulink, responsible for performing different types of gestures and eyes color, was deployed on the Arduino board in advance. Meanwhile, the Raspberry Pi Simulink was working in a real-time manner for controlling and switching among the predefined gestures and eyes color on the Arduino board.

*Social robot*

A low-cost social robot called "Rapiro" was chosen to provide neurofeedback for the subject [57]. The Rapiro robot is a humanoid robot kit with 12 servo motors and an Arduino-compatible controller board. Its capabilities for performing and controlling multiple tasks can be extended by employing a Raspberry Pi board assembled in the head of the robot. For this, the Arduino board in Rapiro could be modified and connected to the Raspberry Pi board. Using a Raspberry Pi board as the brain of robot enabled us to communicate with the robot and send the command signals from the PC to the social robot in real time. Rapiro was selected to provide neurofeedback to the subject by executing movements, playing sounds, and flickering lights in response to specific control commands received via Raspberry Pi Simulink. Rapiro was programmed such that the right target control on monitor (right imaginary movement) would activate the right hand movement of the robot and change its eye color to green; the left target control on monitor (left imaginary movement) would activate the left hand movement of the robot and change the eye color to blue; the top target control (top imaginary movement) led to both hand movements in robot with a mixture of green and blue eye color; and finally the bottom target control (bottom imaginary movement) activated head shaking and altered the eye color to red.

**Results**

For each subject, the EEG data collected during the training protocol were analyzed. As a test example, one subject was randomly selected to perform the brain-robot interaction experiments, during which the EEG data were recorded and reported.

*Training and cursor control task*

For more accurate prediction of imagined body kinematics, we employed a time window of five points (samples) in EEG memory data in our algorithms. Figures 3 and 4 show two sample plots of estimated velocities for one random subject (N2) during horizontal training and vertical training, respectively. These figures illustrate the observed cursor velocities (horizontal and vertical) versus decoded velocities for subject (N2) by using a regression model. Table 1 reports the results for all 5 subjects during the test phase for 1D cursor control tasks. Four subjects each conducted 6 trials of vertical movement and 6 trials of horizontal movement. One subject conducted 6 trials of vertical movement and did not conduct horizontal movements. The total success rate in hitting the targets shows more accuracy in horizontal movement compared to vertical movement. The subjects reported that it was easier to hit the targets in horizontal direction. These results are in consistent with previous noninvasive studies [39, 53, 58, 59].

**Tab. 1:** Results of cursor control experiments.

|  | Vertical Direction | Horizontal Direction |
|---|---|---|
| Number of Trials | 30 | 24 |
| Success Rate (standard deviation) | 83.3% (+/- 11.7%) | 100% (+/- 0%) |

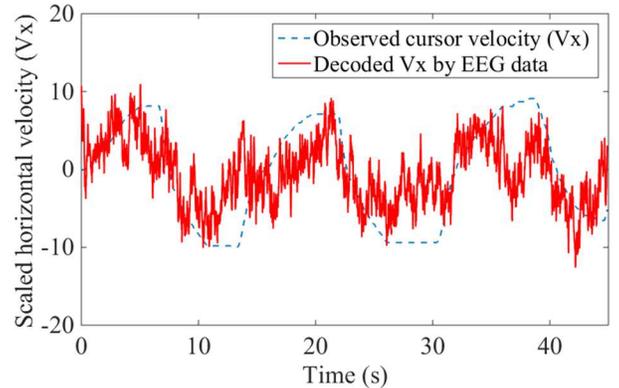

**Fig. 3:** Comparison between observed cursor velocity and decoded cursor velocity in horizontal direction (for subject N2).

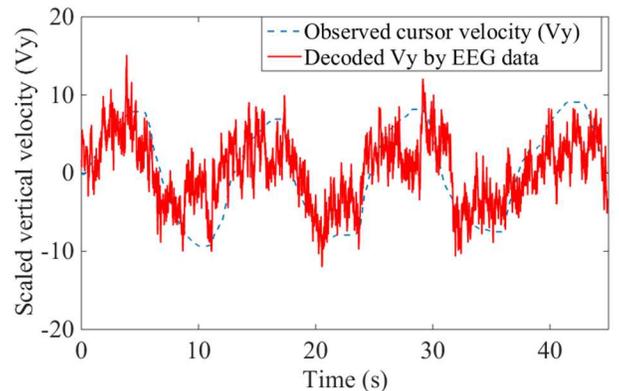

**Fig. 4:** Comparison between observed cursor velocity and decoded cursor velocity in vertical direction (for subject N2).

**Robot control and neurofeedback**

*Offline control*

After performing the test phase in cursor control task by subjects, the data (controlled cursor position) of this phase were collected and applied to control the movements of the social robot in offline mode. Figures 5 and 6 show the series of controlled cursor position data corresponding to the performance of subject (N2) whose training EEG data were shown in Figures 3 and 4, respectively. These position data were sent in offline mode to the Simulink to control the different body parts of social robot (e.g. right hand & green eyes/left hand & blue eyes/two hands & mixture of green and blue eye colors/head shaking & red eyes).

Figure 5 illustrates the cursor position mentally controlled by the subject (N2) during horizontal trials. The center of the screen, where the cursor started to move, is considered as reference point (0, 0). Positive values show the controlled cursor is on the right side of the center and negative values show the cursor is on the left side of the center. After a pre-run time, the trials began and RT (Right Target) or LT (Left Target) showed where the target appeared on screen. The subject had a limited time (15s) to hit the targets or the next trial would begin. In this run, the subject hit all the targets and as it is shown in Figure 5, in all 6 trials the subject moved the cursor to the right side (positive values) for RT and left side (negative values) for LT. The subject hit all the targets in all trials although he struggled a little at the start of trials to guide the cursor in correct direction corresponding to the presented target. The achieved cursor position data were fed to Simulink to control the movements of our social robot. As a simple experiment, it was programmed such that the social robot showed right hand movement & changing eye color to green for positive values (related to RT) and left hand movement & changing eyes color to blue for negative values (related to LT). The robot performed continuous movement of right hand movement & changing the eyes color to green (for RT and compatible with right imaginary movement) or left hand movement & changing the eyes color to blue (for LT compatible with left imaginary movement).

In similar procedure, we fed the data from Figure 6 to Simulink to control some other gestures of the social robot. Figure 6 illustrates the cursor position controlled by the subject (N2) during vertical trials. The center of the screen was again considered as the reference point (0, 0). Positive values show the controlled cursor is above the center and negative values show the cursor is below the center. After a pre-run time, the trials began and TT (Top Target) or BT (Bottom Target) showed where the target appeared on screen. The subject had a limited time (15s) to hit the targets. In this run, the subject hit all the targets by moving the cursor to the top side (positive values) for TT and bottom side (negative values) for BT. But, as it is shown in Figure 6, there are more fluctuations in controlling the cursor to the correct direction compared to horizontal control task in Figure 5. For example, for the first and third trials in Figure 6, the subject guided the cursor to the wrong direction and after some seconds, the subject learned to mentally control it to move in the correct targeted direction. As it is presented in Table 1, controlling the vertical direction (compared to horizontal direction) was more challenging for the subjects. The achieved cursor position data for vertical direction, were fed to Simulink for the movement control of social robot. Here, it was programmed such that the social robot performed two-hands movement & changing eye color to mixture of green & blue for positive values (related to TT) and head shaking & changing eye color to red for negative values (related to BT). The robot executed continuous movement of both hands & changing eye color to mixture of green & blue (for TT and compatible with top imaginary movement) or head shaking & changing eyes color to red (for BT compatible with bottom imaginary movement).

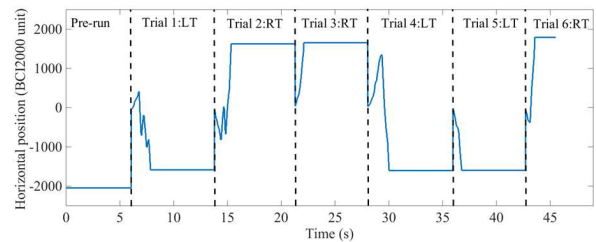

**Fig. 5:** Recorded values of mind-controlled cursor position during one run (6 trials) of cursor control in horizontal direction by a subject (N2). RT: Right Target appeared. LT: Left Target appeared.

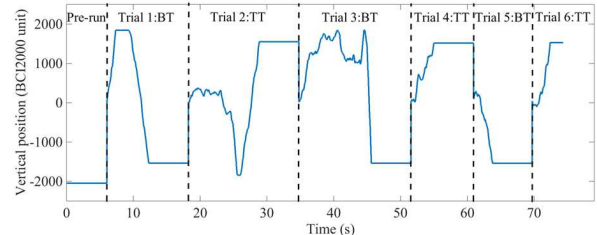

**Fig. 6:** Recorded values of mind-controlled cursor position during one run (6 trials) of cursor control in vertical direction by a subject (N2). TT: Top Target appeared. BT: Bottom Target appeared.

*Online control*

After offline tests, a subject randomly picked from the previously-trained subjects performed online tests and controlled the gestures of robot and eye color in real-time interaction with BMI platform and simultaneously received the corresponding neurofeedback from the robot. As shown in Figure 2, the controlled signals were sent to Simulink through UDP protocol and Simulink controlled the robot based on desired commands. Here, for online testing, we switched to employ another kinematic parameter (velocity) to control the robot in real time. In other words, the mentally- controlled cursor velocities were sent to Simulink and were analyzed to control the robot's movements based on the controlled cursor direction response to the presented target. If the cursor direction were in the correct direction (to the direction of presented target), the predefined movement was activated. Otherwise, no movement occurred.

Figure 7 shows the results of online performance of the subject during horizontal direction control of cursor (1D) and simultaneous control of the robot movements in real time. It illustrates the results of 10 trials of showing targets on right and left sides of screen. The targets were presented to the subject in random manner and the subject could control the social robot by performing right imaginary movement and left imaginary movement. It was programmed such that right imaginary movement and left imaginary movement were mapped to right hand movement & green eyes and left hand movement & blue eyes for the robot, respectively. Figure 7 shows that the subject could activate the robot movements corresponding to the cursor control task. In some trials, for example, the first and second trials, the subject achieved hitting the target in less time (thinner bar width) compared to the other trials. Also, in some trials there are discontinuities in the bar plots as the result of guiding the cursor in wrong directions by the subject.

In a similar way, the vertical control of cursor task (1D) was mapped to control two hands (right & left) movement and head shaking of the robot. Figure 8 shows the results of real-time performance of the subject during vertical direction control of cursor and simultaneously control the robot movements in online mode. Similarly, it illustrates the results of 10 trials of showing targets on top and bottom portions of the screen. The targets were presented to the subject in random manner. By performing top imaginary movement and bottom imaginary movement, the robot was controlled by the subject. It was programmed so that top imaginary movement and bottom imaginary movement were mapped to two-hands movement & mixture of green and blue colors for eyes and head shaking & red eyes for the robot, respectively. As shown in Figure 8, the subject hit the targets on top and bottom and simultaneously activated the corresponding movements in the social robot. As it was discussed based on Table 1, the subject found it harder to control the vertical direction of cursor, particularly for the bottom targets. Figure 8 confirms that the movements are more in activation status compared to horizontal task since it took more time to hit the targets. As it can be seen in most of these trials, the duration of movements (bar width) are longer than those trials seen in Figure 7 and also the similar discontinuities in activation are caused by guiding the cursor in wrong direction.

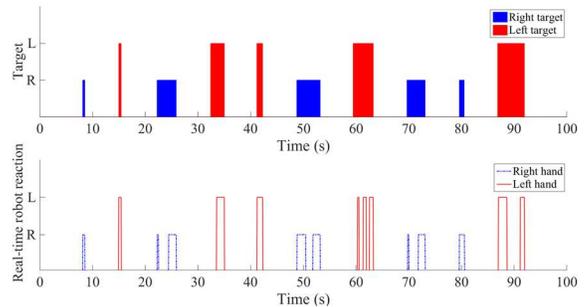

**Fig. 7:** Real-time mind control of robot (right hand movement & green eyes/left hand movement & blue eyes) based on cursor control task performance in horizontal direction during 10 trials for the trained subject. It was programmed such that mental control of the right target control caused right hand movement & green eyes for the robot and left target control caused left hand movement and blue eyes for the robot, which served as simultaneous neurofeedback to the human subject.

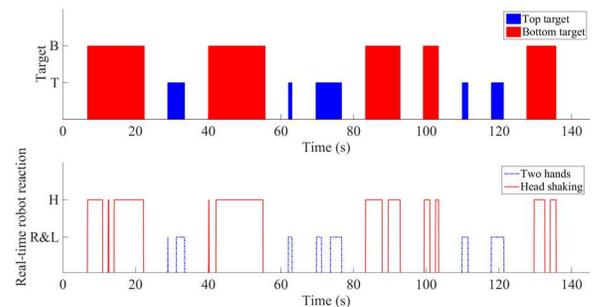

**Fig. 8:** Real-time mental control of robot (two-hand movement & mixture of green and blue colors for eyes / head shaking & red eyes) based on cursor-controlled task performance in vertical directions during 10 trials for the trained subject. The robot movement was programmed to respond to mind-intended directions, i.e., the top target control led to two-hand movement and mixture of green and blue eyes colors; bottom target control led to head shaking & red eyes. The robot's movements and eye-color changes served as simultaneous neurofeedback to human subject.

To make the task more complicated and involved with more neurofeedback from the social robot, the decoder was activated to control the cursor in all areas of the screen by the subject's mind. The subject was asked to control the cursor in 2D space by the same imaginary movements defined for horizontal and vertical tasks and hit the targets which randomly appeared on right, left, top and bottom sides of screen. Meanwhile, it was programmed such that if the subject guided the cursor in correct direction toward the shown target, the robot would show the predefined gesture for that specific target. Figure 9 illustrates the results of cursor task control and activation of four different gestures and eyes colors for the robot as they were defined and explained in horizontal and vertical control tasks in Figures 7 and 8. This task was more challenging for the subject as it is shown in Figure 9. For example, it was hard for the subject to guide the cursor toward the bottom targets and the movement time is longer for this target in comparison to other movement time for the other targets. Also, due to directional errors in the guidance of the cursor by the subject in some inter-trial times, discontinuities were seen in the activation of robot movements.

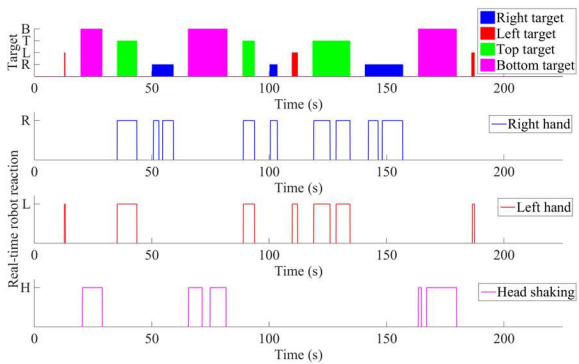

**Fig. 9:** Real-time mind control of robot (right hand movement & green eyes/left hand movement & blue eyes/two hands movement & mixture of green and blue colors for eyes / head shaking & red eyes) based on cursor control task performance in horizontal and vertical directions during 12 trials for the trained subject. The generated neurofeedback movements for specific target is similar to those movements defined in Figures 7 and 8. The robot's movements and eye-color changes served as simultaneous neurofeedback to human subject. The right target control would activate the right hand movement & green eyes; the left target control would activate the left hand movement & blue eyes; the top target control lead to both-hand movement & mixture of green and blue colors for eyes; and the bottom target control activated head shaking & red eyes.

**Discussion and conclusions**

The current study explored a new neuro-based BCI robotic platform using a personalized social robot to give neurofeedback during brain training. Recently, fMRI-based neurofeedback systems have been investigated for cognitive rehabilitation [60]. EEG signals have much higher temporal resolution compared to fMRI methods that are also expensive with poor portability. In contrast, real-time EEG-based neurofeedback systems are much more cost effective and highly portable [61]. In early neurofeedback systems, the human subjects directly interacted with a computer and a screen to receive real-time sensory neurofeedback corresponding to their performances. The advance in the human-robot interaction field in the past decade has shown great promise in rehabilitation programs for patients with brain disorders. In contrast to traditional computer-based feedback system, the interaction with a social robot can provide a more interactive type of communication which encourages active user participation more effectively in cognitive rehabilitation.

Several BMI robotic platforms have been developed for patients with motor disorders for restoring cortical plasticity underlying different movements of human body. Here, as a pilot study, we demonstrate a novel neurofeedback-based BCI platform as a testbed for cognitive training in patients with cognitive deficits. A real-time and EEG-based BMI system was integrated with a social robot to provide neurofeedback to the subject. For initial testing of platform, a new EEG paradigm based on continuous decoding of imagined body kinematics was used. First, the subjects were instructed through a brief training protocol to control a computer cursor on a monitor. Then, the trained subjects were allowed to interact with a BMI platform to control the different gestures and eyes color of a social robot and received neurofeedback from the robot. The online results verified the good performance of the subject in controlling and receiving neurofeedback from the robot's movements and eye color changes; it also confirmed that control with vertical imaginary movements was less accurate just as it was shown during the cursor control task. The work here serves as a feasibility study for the application of the platform for possible development and testing with cognitive algorithms of patients [24, 62-64]. As the next step, we will integrate the current platform with neurofeedback during actual cognitive tasks to improve effectiveness of brain training in patients having cognitive disorders.

**Acknowledgements**

This work was in part supported by a NeuroNET seed grant to XZ; and in part by the NIH under grants NIH P30 AG028383 to the UK Sanders-Brown Center on Aging, and NIH NCRR UL1TR000117 to the UK Center for Clinical and Translational Science. The authors are grateful for useful discussions of Dr. Nancy Munro.